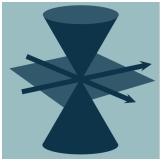
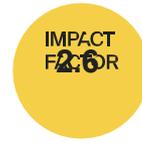
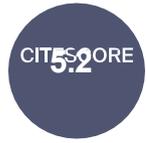

# universe

**Review**

# Detectors and Shieldings: Past and Future at LUNA

Chemseddine Ananna, Lucia Barbieri, Axel Boeltzig, Matteo Campostrini, Fausto Casaburo, Alessandro Compagnucci, Laszlo Csedreki, Riccardo Maria Gesue, Jordan Marsh, Daniela Mercogliano et al.

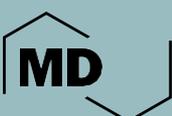
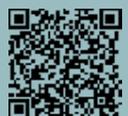

# Detectors and Shieldings: Past and Future at LUNA

Chemseddine Ananna [1], Lucia Barbieri [2], Axel Boeltzig [3], Matteo Campostrini [4], Fausto Casaburo [5], Alessandro Compagnucci [6], Laszlo Csedreki [7], Riccardo Maria Gesue [6], Jordan Marsh [2], Daniela Mercogliano [1], Denise Piatti [8,*], Duncan Robb [2], Ragandeep Singh Sidhu [2] and Jakub Skowronski [8]

1. Dipartimento di Fisica, Università di Napoli Federico II and INFN, Sezione di Napoli, Strada Comunale Cinthia, 80126 Napoli, Italy
2. SUPA, School of Physics and Astronomy, University of Edinburgh, Edinburgh EH9 3FD, UK
3. Helmholtz-Zentrum Dresden-Rossendorf, Bautzner Landstr. 400, 01328 Dresden, Germany
4. INFN Laboratori Nazionali di Legnaro, Via dell'Università 2, 35020 Legnaro, Italy
5. Dipartimento di Fisica, Università degli Studi di Genova and INFN, Sezione di Genova, Via Dodecaneso 33, 16146 Genova, Italy
6. Gran Sasso Science Institute, INFN, Viale Francesco Crispi 7, 67100 L'Aquila, Italy
7. HUN-REN Institute for Nuclear Research (HUN-REN ATOMKI), P.O. Box 51, H-4001 Debrecen, Hungary; csedreki@atomki.hu
8. Dipartimento di Fisica e Astronomia "G. Galilei", Università degli Studi di Padova and INFN, Sezione di Padova, Via Francesco Marzolo 8, 35131 Padova, Italy; jakub.skowronski@pd.infn.it
* Correspondence: denise.piatti@pd.infn.it

**Abstract:** Nuclear reactions are responsible for the chemical evolution of stars, galaxies and the Universe. Unfortunately, at temperatures of interest for nuclear astrophysics, the cross-sections of the thermonuclear reactions are in the pico- femto-barn range and thus measuring them in the laboratory is extremely challenging. In this framework, major steps forward were made with the advent of underground nuclear astrophysics, pioneered by the Laboratory for Underground Nuclear Astrophysics (LUNA). The cosmic background reduction by several orders of magnitude obtained at LUNA, however, needs to be combined with high-performance detectors and dedicated shieldings to obtain the required sensitivity. In the present paper, we report on the recent and future detector- shielding designs at LUNA.

## 1. Introduction

Nuclear reactions shape the life and death of stars and produce most of the chemical elements in the Universe. To improve our knowledge of stellar chemical evolution, precise measurements of the cross-section of nuclear reactions involved are required. The charged particle-induced reaction cross-sections at stellar energies are as small as pb and fb, ham- pering any direct measurements in accelerator-based laboratories at the Earth's surface, where the signal-to-background ratio is too low, mainly because of the influence of cosmic rays. In the absence of experimental data, a way to obtain cross-section information at astrophysical energies is to extrapolate from results attained at higher energies [1]. Such ex- trapolations can introduce significant uncertainties in the cross-section. An example of such uncertainty would be the unknown contribution from a narrow or a large sub-threshold resonance or the screening effect [2], leading to large uncertainties of the reaction rates in stellar models. Direct measurements of the cross-section at or as close as possible to the relevant energies are preferable to reduce the reaction rate uncertainties and improve the precision of stellar models. To achieve the required sensitivity for direct measurements, a combination of high detection efficiency, low background rate in the signal region of interest and high-performance accelerators are vital [3].

 

In recent years there has been a significant improvement in measurements of low energy cross-sections thanks to underground facilities, pioneered by the LUNA [4,5] and recently with several new either deep- or shallow-underground accelerator facilities [6–9]. LUNA started its activity in 1991 with a 50 kV electrostatic accelerator [10], now decommissioned, installed at Laboratori Nazionali del Gran Sasso (LNGS) [3], under 1400 m of rock which acts as a natural shield against cosmic rays and ensures an ultra-low background environment. The background induced by cosmic rays is reduced by six orders of magnitude for muon-induced background and by three orders of magnitude for neutron-induced reactions in $\gamma$ spectra [11]. Likewise, detector setups for charged particles have been observed to benefit from lower background rates at an underground location [12]. Moreover, passive shielding around the detector is more effective than on the surface, where secondary radiation produced by cosmic ray interactions with the shielding itself acts as a further source of background.

In 2001 the LUNA collaboration installed a new high-performance accelerator, LUNA400 kV [13], which is still active today, aimed at the investigation of key hydrogen burning and Big Bang Nucleosynthesis (BBN) reactions. The LUNA400 kV provides $H^+$ and $He^+$ beams to two beamlines, one hosting an extended gas target and the other a solid target station. LUNA is now entering a new phase focused on helium and carbon burning, with the installation at the LNGS Bellotti-Ion Beam Facility (B-IBF) of a 3.5 MV Singletron accelerator, which provides intense proton, alpha and carbon beams [14].

The last ingredient to increase sensitivity, the detection efficiency depends on the detector type and size in addition to the detector-target geometry and materials in between. Complementary detector systems have been used for past cross-section measurements at LUNA, and have often been combined with dedicated passive shielding to further reduce the residual background to achieve the goal sensitivity for the specific scientific case under investigation.

In this article, we present recent improvements in the experimental setups in use at LUNA which make use of different detection systems and shielding to increase the sensitivity to reactions with very low cross-sections. An introduction to the Monte Carlo simulation tools is crucial for both the design and the characterization phase of the detection setups and is given in Section 2. Sections 3–5 detail the recent improvements in shielding and detection setup at LUNA, the resulting enhancement in sensitivity and future developments for $\gamma$, $\alpha$ and $n$ detectors.

## 2. Simulations and SimLuna

Monte Carlo simulations are a useful tool for both setup development and characterization. In this regard, the LUNA collaboration has extensively used Monte Carlo codes to predict the performances of very massive shielding, characterize the response of detectors in specific geometries, and more in general, develop suitable setups for extremely low statistics experiments [15]. Recently, the LUNA collaboration has developed SimLUNA, a GEometry ANd Tracking 4 (GEANT4)-based framework [16] that allows for easy access and updates to the reaction database and setup geometries.

SimLUNA is a multi-platform utility that eases the implementation of the full experimental setup in a GEANT4 simulation. Its organization is modular so that common parts of the setup and physics can be re-used and expanded to recreate different experiment geometries, and it is flexible enough to allow for the definition of experiment-specific elements. Most of the features in the setup (e.g., beam features, detector position, data storage, etc.) can be defined by the user using macro commands, significantly reducing the time needed for testing different configurations of a particular setup.

The relevant features of simulated events are stored within ROOT files [17] which facilitates extraction of key parameters required for the characterization of an experimental setup such as energy loss, straggling, and detection efficiency.

The simulations produced within the SimLUNA framework were validated using extensive data on radioactive sources and nuclear reactions acquired by the LUNA col-laboration. These data were also used to optimize the inelastic reaction model for low energies and expand the energy level and cross-section database (e.g., G4PARTICLEXS and PhotonEvaporation).

Once the simulation for a specific setup is set and validated this is used to precisely estimate detector efficiencies (down to 3% uncertainty), true coincidence summing effects and straggling corrections. The well-tested modules of SimLUNA, coupled with the GEANT4 tools, allow us to design and predict new apparatus sensitivity and performance.

In the future, neutron emission processes will be implemented in SimLUNA in order to account for the new experiments being performed and planned by LUNA collaboration.

## 3. γ Detectors

Since LUNA's advent, the cross-sections of key reactions relevant to many stellar sce- narios have been measured, providing major improvements in our understanding of stellar nucleosynthesis and ultimately of the chemical evolution of the Universe. Each scientific case investigated at LUNA has required specific experimental setups in order to achieve the proper sensitivity. In the following sections, we review the experimental setup recently used at LUNA and the future follow-up, focusing on detector-shielding combinations.

Many of the most important reactions in both stellar hydrogen burning and Big Bang nucleosynthesis proceed either solely or with a high probability, by the emission of γ rays. Over the years, the LUNA collaboration has employed several types of γ detectors to measure some of these reactions [18,19] at low bombarding energies.

### 3.1. The $^{20,21,22}$Ne(p, γ)$^{21,22,23}$Na Reactions

At LUNA, High Purity Germanium (HPGe) detectors are the primary semiconductor detectors utilized for γ detection. These are widely used because of their higher energy resolution, typically around 0.1 % at γ energies near 1 MeV, when compared to other detectors, for example, scintillator detectors. Their high atomic number ($Z$ = 32) ensures a good absorption coefficient. However, these detectors have poor detection efficiency. Hence these detectors are often used when either the expected count rate is high, or when the resolution is needed to resolve nearby γ peaks from each other.

In the recent past, HPGe detectors have been used in both beamlines of the LUNA 400 kV accelerator. In the gas target beamline, a detection system based on two HPGe detec- tors has been widely and successfully utilized [18,20,21], see Figure 1. The detectors are a CANBERRA low background detector with relative efficiency of 130%, and an ORTEC low background detector with relative efficiency of 90%. Both detectors are shielded by a few centimeters of high-conductivity oxygen-free copper and about 25 cm of low-radioactivity lead. The copper layer absorbs the bremmsthralung γ-rays from the beta particles from the decay of $^{210}$Bi, the daughter of $^{210}$Pb. An anti-radon box, made of Plexiglas and flooded with nitrogen gas, is added to remove any residual radon contamination [18,20,21].

The multi-layer shielding is crucial to reduce the environmental background, by three orders of magnitude, at $E_\gamma \leq 3$ MeV, see Figure 2, where γ-rays of interest are found for the $^{20,22}$Ne(p, γ)$^{21,23}$Na reactions and more recently for the $^{21}$Ne(p, γ)$^{22}$Na reaction campaign. The latter is of crucial importance to improve the present understanding of novae, since the detection of the $^{22}$Na decay (half-life of 2.602 years) is one of the main evidence of an underlying O-Ne white dwarf [22]. Moreover, a group of meteoritic material shows a strong excess in $^{22}$Ne, the daughter nucleus of the unstable $^{22}$Na, which is believed to probe NeNa activation during core-collapse supernovae [23]. In both scenarios the observed abundance of $^{22}$Na is not reproduced by models because of the high uncertainty of the $^{21}$Ne(p, γ)$^{22}$Na reaction rate [24].

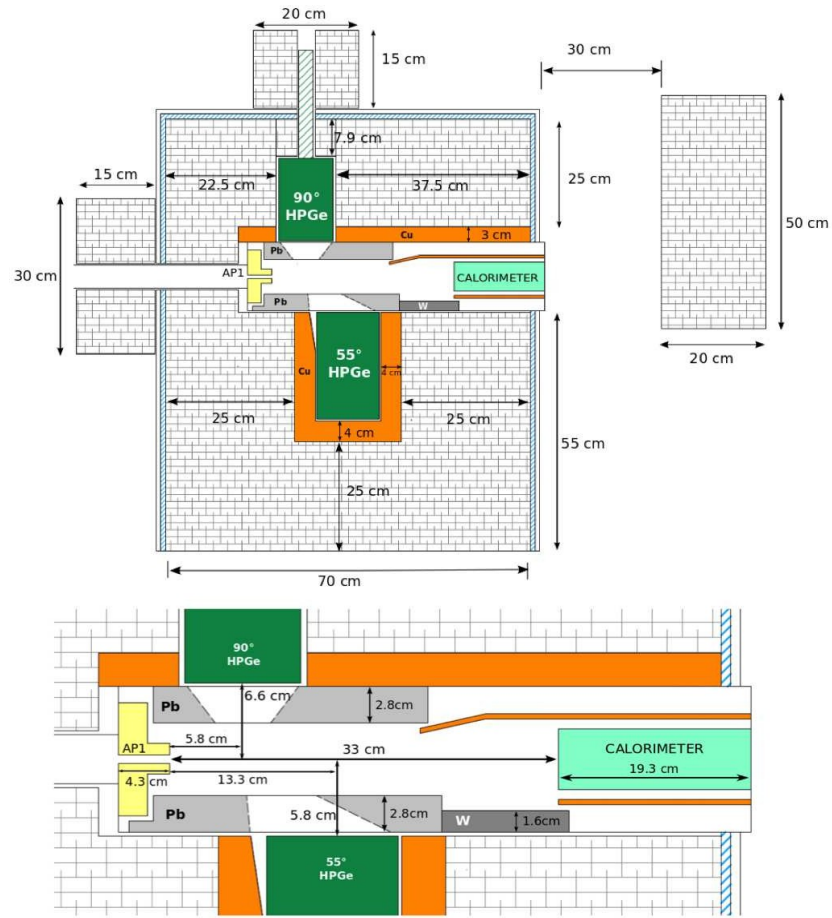

**Figure 1.** (Adapted from [25]), scheme of the setup used in the $^{22}$Ne + $p$ reaction campaign.

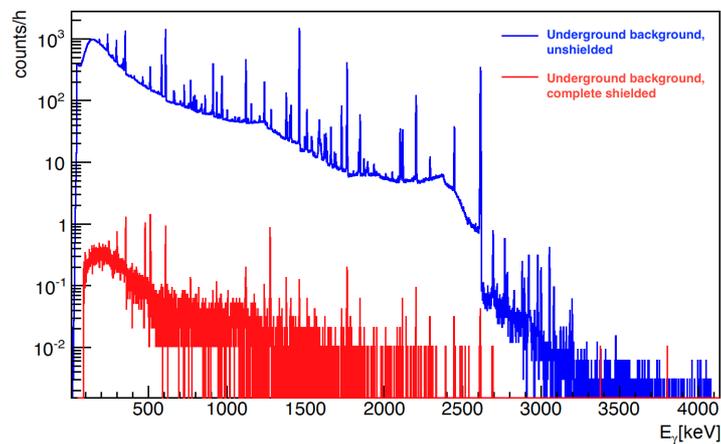

**Figure 2.** (Adapted from [26]), comparison between a background spectrum taken without the shielding, in blue, and with the complete shielding, in red. The two spectra have been scaled for the same live time.

The dominant resonances are at the proton beam energies of $E_p$ = 126 and 272 keV in the laboratory system. Additional resonances at $E_p$ = 271, 290, and 352 keV contribute to the reaction rate <5% in the astrophysical temperature range of interest. All the aforemen- tioned resonances have been measured at LUNA and thanks to the sensitivity achieved with the described setup, high-precision results will soon be published.

*3.2. The* $^2$H(p, γ)$^3$He *Reaction*

An unshielded HPGe detector, installed at the LUNA400 kV gas target, was used for the ambitious goal of achieving a total accuracy of 3% on the S-factor of the $^2$H(p, γ)$^3$He re- action [19]. Such a high precision was required to compare the BBN outputs and deuterium abundance observations and eventually independently probe the baryon density [27,28]. The available data sets were affected by systematic errors of 9% or higher [29–32]. Moreover, the ab initio calculation by [33] disagreed at the 8% level with the fit of experimental data by Iliadis et al. [34].

The $^2$H(p, γ)$^3$He reaction proceeds via a direct capture mechanism emitting single γ-rays with energy between 5.5 and 5.8 MeV, where we could exploit the six orders of mag- nitude suppression of cosmic background at LUNA. The γ-rays emitted by the $^2$H(p, γ)$^3$He reaction were detected by a large HPGe at 90° with respect to the beam axis. In order to
achieve a total accuracy of 3 % on the *S*-factor, it was crucial to minimize all the sources
of systematic uncertainties. The γ-ray detection efficiency $\epsilon(z, E\gamma)$ was carefully evaluated as a function of both the γ-ray energy and emission point in the target chamber [35] by using radioactive sources and exploiting the $E_{cm}$ = 259 keV resonance of the $^{14}$N(p, γ)$^{15}$O reaction [35]. The interaction between the proton beam and deuterium gas can take place at different positions along the beam axis, causing the emission of photons subtending different angles with respect to the HPGe detector. The efficiency was measured at different target positions with radioactive sources, $^{60}$Co and $^{137}$Cs, and exploiting the $E_{cm}$ = 259 keV resonance of the $^{14}$N(p,γ)$^{15}$O reaction. This resonance produces γ-ray cascades over a wide energy range (0.7–7 MeV), allowing to extend the efficiency measurement to cover the $^2$H(p, γ)$^3$He region of interest [35]. For the case of the $^{14}$N(p, γ)$^{15}$O reaction, a second HPGe detector was added to the setup. This additional detector could move along the gas target. Thanks to a slit collimator mounted in front of the detector and dedicated lead shielding the detected γ-rays are coming from the resonance excited in front of the detector. In this way, it can be used to gate the detector used for the $^2$H(p, γ)$^3$He reaction measure- ment, see Figure 3. This setup design together with well-constrained simulation allowed us to obtain a detection efficiency uncertainty of 2 % at most, see Figure 4. More details on the determination of the aforementioned quantities and uncertainties are available in [35].

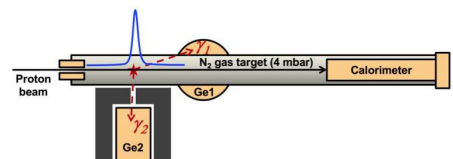

**Figure 3.** (Adapted from [35]), scheme of the setup used for the accurate efficiency measurement in the framework of the $^2$H(p, γ)$^3$He reaction cross-section measurement.

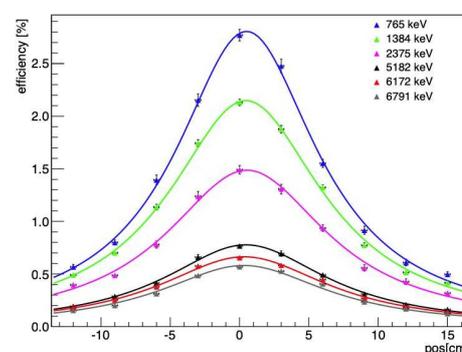

**Figure 4.** (Adapted from [35]). Photo-peak efficiency measured for six γ-ray energies as a function of source position along the beam axis (z = 0 corresponds to the center of the target chamber). Errors are statistical only. Curves represent simulated efficiencies for point-like sources.

*3.3. The $^{17}$O(p, γ)$^{18}$F Reaction*

At LUNA a Bismuth germanate oxide (BGO) detector has also been used for its high efficiency, close to 50% for a photopeak at 10 MeV [36] in the addback spectrum, and a nearly 4π solid angle coverage. Such a gain in efficiency allows experimental campaigns to measure reactions at very low bombarding energies where the cross-section is extremely low. The LUNA 4πBGO detector consists of six independent crystals and has been mainly used in the Total Absorption Mode, namely in the offline analysis the add-back spectrum is reconstructed with a coincidence window of 3.5 μs (cf. [37]). This puts the sum energy γ peak at high energies, generally above the threshold of the detector's internal background, exploiting the energy region where the cosmic background is already reduced by the LNGS environment. However, the sensitivity required for the recent investigation of the tentative resonances at $E_p$ = 68 and 395 keV in $^{17}$O(p, γ)$^{18}$F and in the $^{22}$Ne(α, γ)$^{26}$Mg, respectively, demanded dedicated shielding to further reduce the residual background. The 68 keV resonance dominates the $^{17}$O(p, γ)$^{18}$F reaction rate at the temperature of interest for the hydrogen-shell burning in giant stars [37]. Oxygen isotopes are observed in giants' atmospheres and in meteoritic grains and are a useful tool to trace nucleosynthesis in these stars, the mixing episodes in the envelope and finally the puzzle of the origin of different grains [38,39]. The 68 keV resonance has so far only been studied indirectly in the (p, γ) channel because of the low count rate [40,41]. A recent campaign at LUNA aimed to directly measure this resonance with a dedicated high sensitivity setup, which combined low absorption material, the efficient 4π BGO and three-layer shielding, composed of (from inner to outer) 1 cm of borated (5%) polyethylene 10 cm of lead and 5 cm of borated (5%) polyethylene, see Figure 5. The designed setup increased the efficiency by 20% with respect to a 10 cm lead shield only and the background was reduced by a factor of 4 in the region of interest, see Figure 6. Details on the setup are available in [15] and in Figure 5 while results will be published soon.

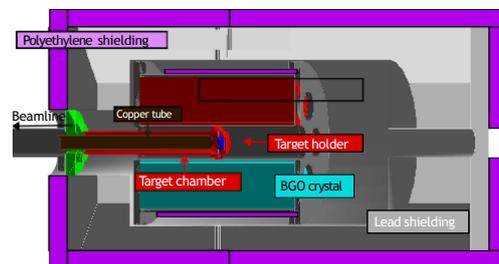

**Figure 5.** (Adapted from [36]), scheme of the setup used in the $^{17}$O(p, γ)$^{18}$F reaction campaign.

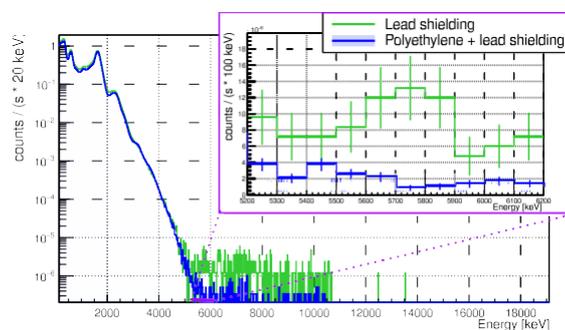

**Figure 6.** (Adapted from [36]), comparison between a background spectrum taken with solely lead shielding, in green, and with the complete shielding, in blue. The two spectra have been scaled for the same live time.

In [15] an alternative technique, which exploits the BGO segmentation, is presented. It was applied in the recently measured $^{12}$C(p, γ)$^{13}$N reaction [15], where the $^{13}$N decays by positron emission. The geometrical proprieties of the 511 keV γ-rays emission after annihilation help to disentangle the intrinsic background from the reaction by looking at the coincidences only in the opposite detectors.

*3.4. The $^{22}$Ne($α$, n)$^{25}$Mg Reaction*

The $^{22}$Ne($α$, $γ$)$^{26}$Mg reaction is the main competitor of the $^{22}$Ne($α$, n)$^{25}$Mg reaction, which is responsible for neutron production for the s-process in massive stars. The 395 keV resonance dominates the rate of the former reaction at temperatures of interest for core helium burning, with this being the main source of uncertainty for the determination of the temperature at which the neutron channel dominates the $γ$ channel. LUNA investigated this resonance by means of the gas target surrounded by the high-efficiency BGO detectors. A first campaign revealed the need to reduce the residual background, mainly neutron-induced background, in the region of interest; the BGO was surrounded by 10 cm thick borated (5%) polyethylene shielding, with a reduction in the background by a factor 3, see Figure 7. This allowed us to derive one of the lowest upper limits ever reported by LUNA [42].

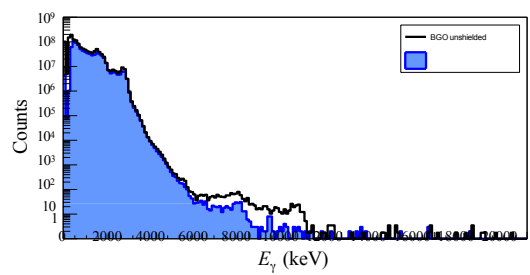

**Figure 7.** (Adapted from [42]), comparison between a background spectrum taken without shielding, in black, and with the borated polyethylene shielding, in blue. The two spectra have been scaled for the same live time.

An alternative technique that exploits the BGO segmentation is the activation technique. It was applied in the recently measured $^{12}$C(p, $γ$)$^{13}$N reaction [15], where the $^{13}$N decays by positron emission. The geometrical proprieties of the 511 keV $γ$-rays emission after annihilation help to disentangle the intrinsic background from the reaction by looking at the coincidences only in the opposite detectors.

*3.5. The Case of the $^{16}$O(p, $γ$)$^{17}$F Reaction*

For some reactions, it is advisable to use both scintillator and semiconductor detectors together to take advantage of the strengths of each simultaneously. An example is the 2022–2023 campaign to measure the $^{16}$O(p, $γ$)$^{17}$F reaction at the LUNA 400 kV accelerator. This reaction is crucial in modeling the evolution of Asymptotic Giant Branch stars (AGB) [43], in particular, the relative abundances of the oxygen isotopes produced in these stars. An ORTEC HPGe with a relative efficiency of 121% was installed at 55°. The high resolution of the HPGe was vital to ensure that the secondary $γ$ from the de-excitation of the 495 keV first excited state in $^{17}$F could be resolved from the 511 keV annihilation peak. The HPGe was supplemented with two 57 mm diameter Scionix CeBr$_3$ scintillator detectors. While the efficiency of these detectors is low compared to the BGO discussed above, it is higher than the HPGe. This was important, as the $^{16}$O(p, $γ$)$^{17}$F reaction is non-resonant, and so the count rate at the low bombarding energies of the LUNA 400 kV accelerator is very low. The three detectors were positioned in close geometry to the Ta$_2$O$_5$ target at about 20 mm from the target, contained within thick lead and polyethylene shields. A future paper by the LUNA collaboration will describe the setup and measurements in detail.

*3.6. Outlooks: The $^{12}$C + $^{12}$C Reaction*

One of the key measurements in the LUNA program will take place at the new B-IBF and aims at measuring the $^{12}$C + $^{12}$C reaction cross-section. This reaction is of crucial importance to determine the evolution of stars, having a significant impact on the $M_{\rm up}$ parameter [44]. While the energies of astrophysical interest are between 1 and 2 MeV, the available direct data for the main fusion channels the $^{12}$C($^{12}$C, $\alpha$)$^{20}$Ne and the $^{12}$C($^{12}$C, $p$)$^{23}$Na reactions, extend down to 2.2 MeV [45–51]. However, a recent indirect measurement reported data down to low energies showing several resonances [52,53], which must be confirmed independently. LUNA is going to directly access the $^{12}$C + $^{12}$C reaction cross-section inside the Gamow window for the first time. The LUNA measure- ment will focus on γ-rays emitted due to the de-excitation of the first excited states in $^{20}$Ne and $^{23}$Na, respectively. The detection system consists of a HPGe detector, with a relative efficiency of 150%, located at 0° with respect to the beam direction. The HPGe detector will be in close geometry, namely at about 20 mm from the target to maximize the solid angle, mitigating the possible angular distribution effect. In addition, an anti-Compton array of NaI scintillators will be installed all around the target and the HPGe detector. The detectors and the scattering chamber will be embedded in dedicated shielding. Different types of targets are now under testing. Preliminary characterization of the setup and measurement at high energies will start soon. LUNA is also working on future phases dedicated to the study of $^{12}$C($^{12}$C, $\alpha$)$^{20}$Ne and the $^{12}$C($^{12}$C, $p$)$^{23}$Na channels via particle detection.

**4. Charged Particle Detectors at LUNA**

Charged particle detectors have been utilized by the LUNA collaboration throughout its history to study key reactions of hydrogen burning via both the pp-chain and the CNO cycles [12,54–56]. Due to their high detection efficiency, small size, and versatility, silicon detectors have been used in charged particle detection experimental setups at LUNA.

*4.1. The Case of the $^3$He($^3$He, 2p)$^4$He Reaction*

The first charged particle detection setup was constructed to study the $^3$He($^3$He, 2p)$^4$He reaction at solar energies (16 keV–27 keV), which was the reason why LUNA was born in 1991. Two different detection setups were used. One setup, made of four ΔE (140 µm thick, 2500 mm$^2$ active area)-E(1000 µm thick, 2500 mm$^2$ active area) telescopes was used for the cross-section measurement down to 20.8 keV, due to beam-induced background limitations [54]. A new and improved setup was used to measure the cross-section down to 16.5 keV. This setup consisted of eight silicon detectors (1 mm thick, 2500 mm$^2$ active area) arranged around the gas target. A 1 µm mylar foil, a 1 µm aluminum foil and a 10 µm nickel cylinder were placed in front of each detector to stop the produced $^4$He nuclei, the intense scattered $^3$He beam and the light induced by the beam [55]. The absorber configuration was crucial to select the events producing two protons, i.e., the $^3$He($^3$He, 2p)$^4$He reaction, against the beam-induced background from the $^3$He($^2$H, p)$^4$He reaction. The $^3$He + $^3$He campaign successfully measured the cross-section down to the lower edge of the solar Gamow peak, ruling out the presence of a resonance that was suggested in the 1990s as a possible nuclear solution to the solar neutrino problem.

*4.2. The $^{17,18}$O(p, $\alpha$) Reactions*

More recently, (p, $\alpha$) reactions have been the primary motivation for the detection of charged particles, with a dedicated experimental setup to study the $^{17,18}$O(p, $\alpha$) reactions in 2015. The objective of these measurements was to reduce uncertainties in the reaction rates that impact the abundance of oxygen isotopes produced in Red Giant Branch (RGB) and AGB stars. The experimental setup, shown in Figure 8, consisted of a scattering chamber made up of two hemispherical domes, an aluminum outer dome and a copper inner dome. The outer dome was used to house eight silicon detectors (five 300 µm and three 700 µm thick detectors). The proton beam entered the chamber from the top of Figure 8, hitting a Ta$_2$O$_5$ target in the center of the chamber. Alpha particles emitted at backward angles were detected by the silicon detectors positioned 6 cm away from the target and at angles of 102.5 and 135 degrees. Aluminized Mylar foils were mounted to the inside of the copper dome to protect the detectors from the high flux of elastically scattered beam protons.

Foil widths were chosen to absorb the low energy protons (<360 keV) but let the higher energy alpha particles (1 to 4 MeV depending on the reaction) through to be detected by the silicon detectors. Stainless steel collimators were also mounted onto the front of each silicon detector to further reduce the flux of higher energy protons. The inner copper dome was negatively biased to suppress secondary electron emission and achieve reliable current readings from the charge deposited on the isolated target.

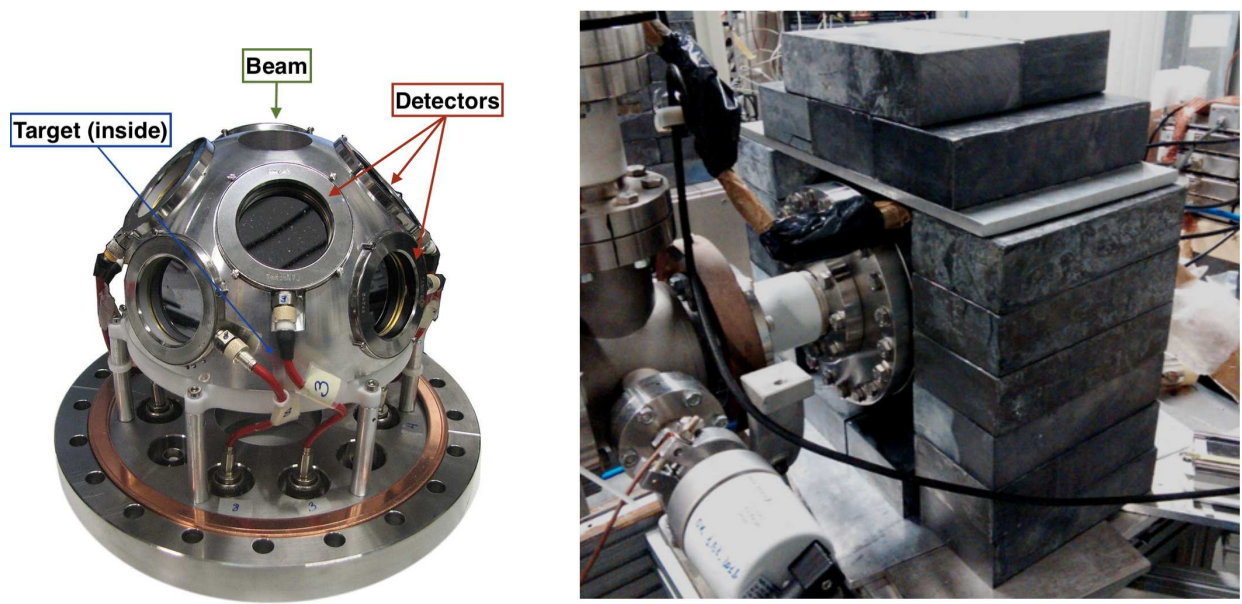

**Figure 8.** (**Left**): (Adapted from [12]) The reaction chamber used in this study. The ion beam enters from the hole in the top, and hits the solid $Ta_2O_5$ target located in the center of the dome (not shown). Alpha particles are detected at backward angles by an array of eight silicon detectors. (**Right**): the lead shielding surrounding the setup, the beam enters from the left.

Measurements focused on the $E_p$ = 70 keV resonance in the $^{17}O(p, \alpha)$ reaction, in addition to the $E_p$ = 95 keV resonance and the general excitation function from ($E_p$ = 60 to 360 keV) of the $^{18}O(p, \alpha)$ reaction [12]. The low cross-section of these reactions made the background suppression from the underground environment of LNGS vital for this experimental campaign. Low energy events below 1 MeV from electrons produced through Compton scattering events and secondary electrons produced through $\gamma$ rays passing through the detectors dominated the background. This background was particularly challenging as the energy region of interest for alpha particles from the 70 keV resonance in $^{17}O(p, \alpha)$ after passing through the Mylar foils was between 200 and 300 keV. The combination of the underground environment and a 5 cm lead shield around the setup resulted in a 14-fold reduction in this background, contributing to the success of this measurement. The background reduction in the underground environment compared to an overground laboratory is shown in Figure 9.

In addition to the background suppression, the calibration of the silicon detectors at low energy was crucial for the success of the campaign. The very low counting rate made accurate determination of the energy of detected events and the region of interest of events from the reaction mandatory to separate events from the background. A two-step process was adopted to calibrate the detectors. First, a pulser walk-through was carried out to determine the ADC offset and then an alpha source and a conversion electron source, in this case, $^{241}$Am ($E_\alpha$ = 5486 keV) and $^{137}$Cs ($E_K$ = 624 keV), were used to complete the calibration of the detector response. To determine the region of interest, the accurate energy loss of the alpha particles in the Mylar foils, which requires accurate determination of the thickness of the protective foils, was needed. Preliminary measurements of the foil thicknesses were conducted by using the energy loss of alpha particles from a mixed triple alpha source. Final calibration of the foil thicknesses and silicon detectors were conducted simultaneously using resonances at $E_p$ = 193 and 151 keV in $^{17}O(p, \alpha)$ and $^{18}O(p, \alpha)$, respectively. An iterative procedure calculated the gain calibration of the detector assuming a certain foil thickness for one alpha peak, then used this calibration to determine the energy of a second alpha peak. This energy was then compared to the expected energy, calculated using SRIM, for the alpha particle if it had passed through the assumed foil thickness. The assumed foil thickness was varied until the difference between the detected energy with the updated gain calibration and the expected energy from the SRIM of the second alpha peak was minimized. Details of this procedure can be found in [12]. This procedure yielded a foil thickness value in agreement with the nominal one (2 μm Mylar + 0.4 μm aluminum coating) from the suppliers, with an estimated inhomogeneity of ±0.1 μm. Thanks to the high sensitivity achieved with the described setup and to the accuracy of the analysis tools developed, LUNA was able to directly measure the resonance strength of $E_p$ = 70 keV $^{17}O(p, \alpha)$ reaction [56] for the first time, with a huge impact on our understanding of the observed abundance of $^{17,18}O$ in AGB stars and in stardust grains [39,57].

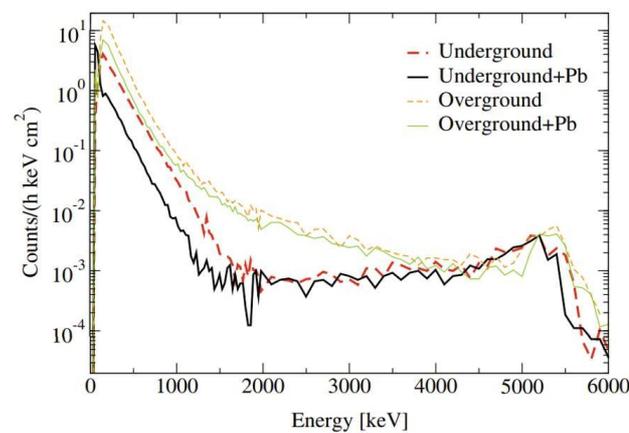

**Figure 9.** Comparison of background spectra taken with and without shielding underground at LUNA and in Edinburgh. The background suppression around 200 keV is up to a factor of 15. The broad peak around 5 MeV is due to intrinsic alpha activity in the silicon detectors [12].

*4.3. Outlooks: The $^{23}Na(p, \alpha)^{20}Ne$ Reaction*

A new LUNA experimental campaign focused primarily on the measurement of the $E_{cm}$ = 133 keV resonance strength in the $^{23}Na(p, \alpha)^{20}Ne$ reaction is scheduled for late March 2024 (ELDAR-burning questions on the origin of the Elements in the Lives and Deaths of stARs, UKRI ERC StG (EP/X019381/1)). This resonance has never been directly observed in overground laboratories, due to the extremely low expected count rate (tenths of counts per day), but will be detectable in the low background environment of LUNA laboratory. A precise estimation of the unknown resonance strength may prove crucial in determining the $^{23}Na(p, \alpha)^{20}Ne$ reaction rate at stellar temperature, with possible impor- tant consequences for globular cluster formation scenarios invoking AGB stars as pristine material polluters and seeds for subsequent stellar generations [58,59]). This scenario has been proposed to reproduce the globular cluster chemical anomalies, such as the observed anti-correlation between sodium and oxygen abundances. Many authors [60,61] have sug- gested that a decrease in the $^{23}Na(p, \alpha)^{20}Ne$ reaction rate by a factor ≈4–5 than currently adopted in [62] could provide a solution to this long-lasting problem.

In this context, following the promising results on (p, α) captures obtained at LUNA, a detailed study has been carried out to design a new silicon detector array. The improved setup will increase the detection efficiency of charged particles and offer the possibility of exploring underlying angular distributions through optimized solid angle coverage. The detection structure has also been designed to host preamplifiers in a vacuum, close to the silicon diodes, minimizing noise pick-up. High-quality diodes offering the possibility of reading both sides of the junction have been selected, including nTDS double-sided PAD detectors, highly suited for applying Pulse Shape Analysis techniques to acquired signals. An innovative design for the inner dome holding the protective aluminized Mylar foil has also been conceived. The silicon array has been installed and is in the last phases of commissioning. Data taking will start in March 2024 and is expected to cover the whole of 2024. The analysis of the results will be carried out utilizing the SimLuna framework, where the geometry of the new detection array is being implemented. This will allow detailed detector efficiency calculations and extensive studies of the interaction of the reaction's alpha particles with the protective foils before their detection. More technical details on the new setup will be the subject of a future paper by the LUNA collaboration.

## 5. Neutron Detectors at LUNA

Charged particle-induced reactions cannot contribute to the synthesis of heavy elements above the iron groups due to the minimum of the nucleon binding energy around Fe and the increment of Coulomb repulsion. Moreover, the production of heavy elements above iron does not release energy, and therefore, it does not contribute to the energy production of stars. Neutron capture on heavy nuclei followed by beta-decay of neutron-rich nuclei is the mechanism that explains most of the observed abundance pattern for $A > 60$. In low neutron density astrophysical sites the so-called astrophysical slow ($s$-)process, i.e., the time scale for neutron capture is longer than the beta-decay lifetime of the newly-created isotope, proceeds along the valley of stability [63–65] and references therein. The helium-burning shell of low-mass AGB stars and the helium-burning core of massive stars are considered the astrophysical site of the main and weak $s$-process components, respectively. The former process is responsible for the production of the $90 \lesssim A \lesssim 209$ isotopes at temperatures around $T_9 = 0.1$. The neutron source for the main $s$-process is the $^{13}$C$(\alpha, n)^{16}$O reaction, $Q$-value 2.216 MeV, whose cross-section must be precisely determined inside the Gamow peak to reliably constrain the nuclear physics input to $s$-process calculations. On the other hand, the $^{22}$Ne$(\alpha, n)^{25}$Mg (Q = −478 keV) nuclear reaction is the main neutron source for the weak component of the $s$-process leading to the nucleosynthesis of elements within the mass range $60 \lesssim A \lesssim 90$. It is active during the He-core and C-shell burning phases in massive stars ($M \geq 8\, M_\odot$). The overall neutron budget available for the nucleosynthesis depends on the $^{22}$Ne$(\alpha, n)^{25}$Mg reaction rate, which presently relies on poorly constrained extrapolation of the cross-section inside the Gamow window [66]. To obtain reliable nucleosynthesis predictions, it is crucial to access the $^{22}$Ne$(\alpha, n)^{25}$Mg cross-section down to low energies. In both cases, the expected count rate is below 1 count/h. The detection setups described in the following aim to increase the neutron detection efficiency, which is crucial to overcoming the environmental and intrinsic backgrounds and estimating the yield.

### 5.1. The $^{13}$C$(\alpha, n)^{16}$O Reaction

The first n-detection experiment at LUNA was designed for the study of the $^{13}$C$(\alpha, n)^{16}$O reaction cross-section [67–69].

Past experiments used multiple approaches to detect neutrons, i.e., high efficiency gas filled proportional counters either based on $^3$He gas [68,70–73] or $^{10}$BF$_3$ gas [74,75]; or high angular resolution scintillators, e.g., [76,77].

In past experiments, the environmental background was a limiting factor. This is made of two main components: the cosmic-ray induced neutron flux [78] and the neutron background by the ($\alpha$,n) reactions induced by $\alpha$ particles emitted by the spontaneous fission of $^{238}$U and $^{232}$Th in the rocks or by the activation of the experimental setup itself. These sources of background can be significantly suppressed by moving the experiment to underground locations and with an accurate selection of the setup materials.

The LUNA neutron array consisted of eighteen $^3$He filled proportional counters with stainless steel housing arranged in two concentric rings around the target chamber.

Moreover, two different geometries (vertical and horizontal) were used to optimize the efficiency in the studied $E_\alpha$ region (as shown in Figure 10 and in more detail in [69]).

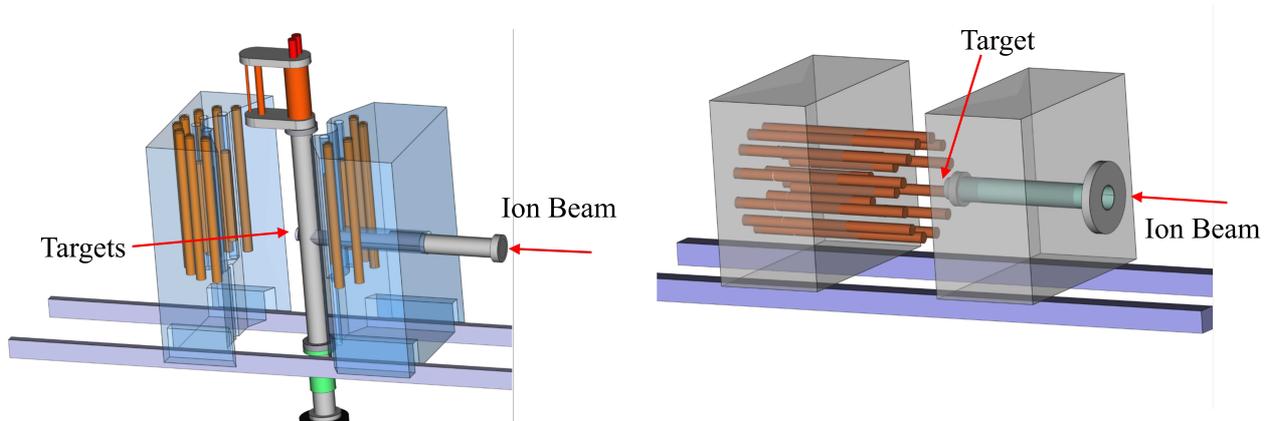

**Figure 10.** Vertical (**left** panel) and horizontal (**right** panel) setup of the LUNA neutron detector array. Orange cylinders and transparent boxes represent the $^3$He neutron counters and polyethylene moderators, respectively. Adapted from [69].

The location of the experimental apparatus of LUNA and the properly selected material of the walls of the $^3$He counters (stainless steel) produced a unique low environmental background with a 3.3 counts/hour counting rate in the detector. The comparison of the detector background in the signal region of interest is shown in Figure 11 in different conditions: data taken in the underground laboratory using a counter made from stainless steel, in the underground laboratory of the LNGS using a single counter made of aluminum, and on the Earth's surface.

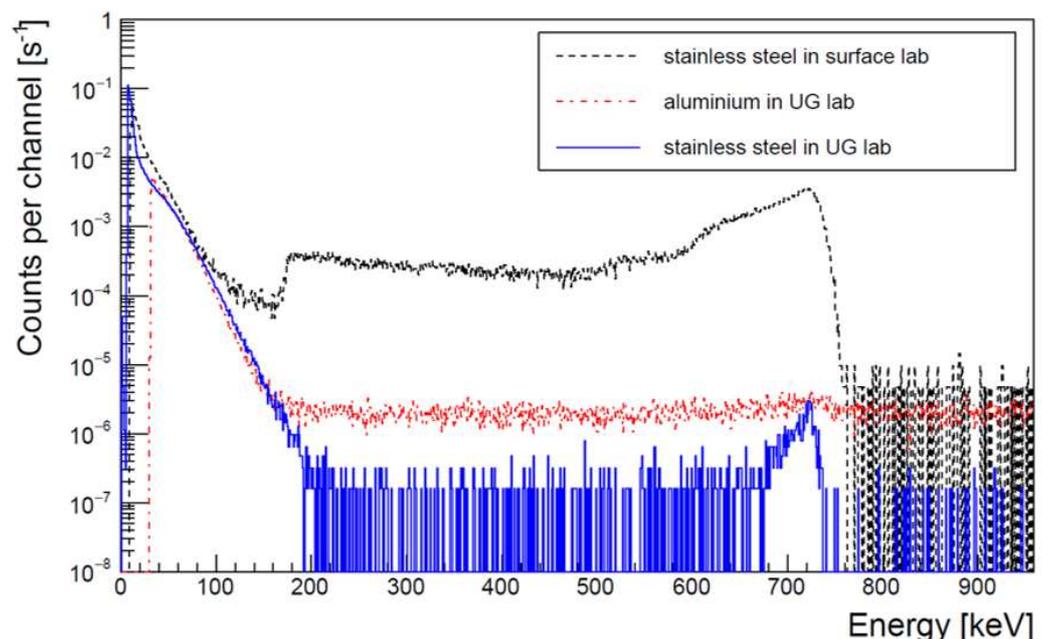

**Figure 11.** Comparison of background spectra of bare $^3$He counters acquired on the earth's surface and in the underground laboratory of LNGS (UG lab) using a single counter made from aluminum (dash-dotted line) and from stainless steel (dashed and solid line). Adapted from [69].

Although a unique low experimental background is achieved, the emitted alphas by the counter walls remained the main contributor to the background. Nevertheless, the detection mechanism of thermal neutrons in $^3$He-counters provides the possibility to discriminate the events triggered by either $\alpha$-s or neutrons. Neutron capture in $^3$He(n, p)$^3$H ($Q$ = 764 keV) is characterized by the emission of two charged particles as a proton (573 keV) and triton (191 keV), thus their energy deposition in the counter can be described by two well-defined Bragg peaks. In the case of $\alpha$-s, a single charge peak follows the energy deposition.

Thanks to the slightly different detection mechanism of neutrons and alphas in the counters, the LUNA collaboration used the Pulse Shape Discrimination (PSD) technique based on digital filtering to convert the integrated signal of the charge-sensitive preampli- fiers to a current pulse. The application of this PSD technique allows the suppression of internal alpha-induced background by >98.5% and reduces the total background of the LUNA neutron array to (1.23 ± 0.12) counts/hour [79,80], which has special importance at the lowest measured energies where the reaction yield of the $^{13}$C($\alpha$, n)$^{16}$O reaction drops
to the 1 event/h level.

Finally, the achieved background rate by LUNA represents an improvement of two orders of magnitude over similar setups [70,72] used in the past and it is a factor of 4 better than the value obtained recently by the JUNA collaboration.

A crucial aspect of the measurement was the efficiency determination. In contrast to γ-ray spectroscopy, the determination of the neutron detection efficiency as a function of neutron energy, $\eta(E_n)$, is challenging mainly due to the limited choices of sources with accurately known energy spectra and/or angular distributions (in the case of nuclear
reactions based sources) and in some cases the limited availability of accurately calibrated sources. To constrain the uncertainty of efficiency determination, the design of the neutron detection setup should be optimized to obtain a detection efficiency as flat as possible along the neutron energy range of interest [81].

The GEANT4 toolkit [82,83] (version 10.03, with "neutron high precision" physics and thermal scattering corrections enabled for water and polyethylene) was used to simulate the detector response of the LUNA neutron array. The simulation was used to calculate the relative distribution of detected events in the counters, the ratio between total yields of the outer and inner rings, and the energy dependence of the neutron detection efficiency. These quantities were compared with experimental data using an AmBe neutron source and the $^{51}$V(p, n)$^{51}$Cr
reaction. Absolute neutron detection efficiency of (37 ± 3)% (horizontal) and
(34 ± 3)% (vertical) of the two setups was obtained in averages in the $E_n$ = 2.2–2.6 MeV range corresponding to the neutron emission of the $^{13}$C($\alpha$, n)$^{16}$O reaction in the LUNA
experiment $\alpha$-energy range. The aforementioned simulation will be soon updated and implemented inside the SimLUNA framework.

Thanks to the low background level and the precise efficiency determination, the LUNA collaboration measured the $^{13}$C($\alpha$,n)$^{16}$O cross-section directly inside the *s*-process Gamow window for the first time, reaching an overall uncertainty lower than 20% and contributing to a better understanding of the evolution of AGB stars and the formation of heavy elements.

*5.2. Outlooks: The $^{22}$Ne($\alpha$, n)$^{25}$Mg Reaction*

The successful campaign of the $^{13}$C($\alpha$, n)$^{16}$O leads the LUNA efforts to move to the other neutron source for the *s*-process: the $^{22}$Ne($\alpha$, n)$^{25}$Mg reaction.

The SHADES (**S**cintillator-**He**3 **A**rray for **D**eep-underground **E**xperiments on the **S**-process) is a neutron detection array designed to perform a direct cross-section mea- surement of the astrophysically relevant $^{22}$Ne($\alpha$, n)$^{25}$Mg nuclear reaction [84]. Due to the negative $Q$-value, neutrons from $^{22}$Ne($\alpha$, n)$^{25}$Mg are characterized by an energy spectrum extending down to almost zero energy, making their experimental detection very chal- lenging. The expected low cross-section also complicates the measurement conditions since the low neutron production rate may become indistinguishable from the intrinsic
background [85]. To improve the setup sensitivity, both the external and beam-induced background need to be reduced. In the SHADES setup, this is achieved by using both passive and active shielding. The SHADES detector array consists of 12 EJ-309 liquid scintillators arranged in a ring and surrounding a recirculating, windowless gas target.

Two other rings of $^3$He proportional counters, composed of 12 units and 6 units, respectively, are placed in between (see Figure 12). The full setup is placed inside a borated polyethylene castle that passively shields it from the external neutron background. The liquid scintillators serve a double function: thermalizing $^{22}$Ne($\alpha$, n)$^{25}$Mg reaction neutrons that did not interact with one of the $^3$He counters, while also giving information about their recoil energy above a certain threshold (a few hundred keV). This latter is made possible thanks to their well-assessed Pulse Shape Discrimination capabilities allowing distinguishing neutron events from γ ones. Neutrons that are thermalized inside one of the scintillators are then scattered and may be captured by a $^3$He counter through the $^3$He(n, p)T (*Q*-value = 764 keV) nuclear reaction characterized by a very high cross-section for thermal neutrons (~100%). This has the direct effect of increasing the total neutron detection efficiency, surpassing any loss due to the active filtering through time coincidence selection. Low-energy neutrons with $E_n <$ 500 keV are also registered by the $^3$He counters with an efficiency of ~15–20%. The $^3$He-EJ309 liquid scintillator hybrid design allows a further suppression of environmental and beam-induced backgrounds through time coincidence vetoing that preliminary analysis has already proven to be well-adapted for this task. Also, conducting the experiment in the deep-underground facility of LNGS using the newly commissioned LUNA-MV accelerator will allow the investigation of resonances located under the current 832 keV state-of-the-art lower limit, the presence of which is suggested by the existence of parity states in this region. New measurements at previously measured energies are also scheduled in order to provide new data comparable to the literature.

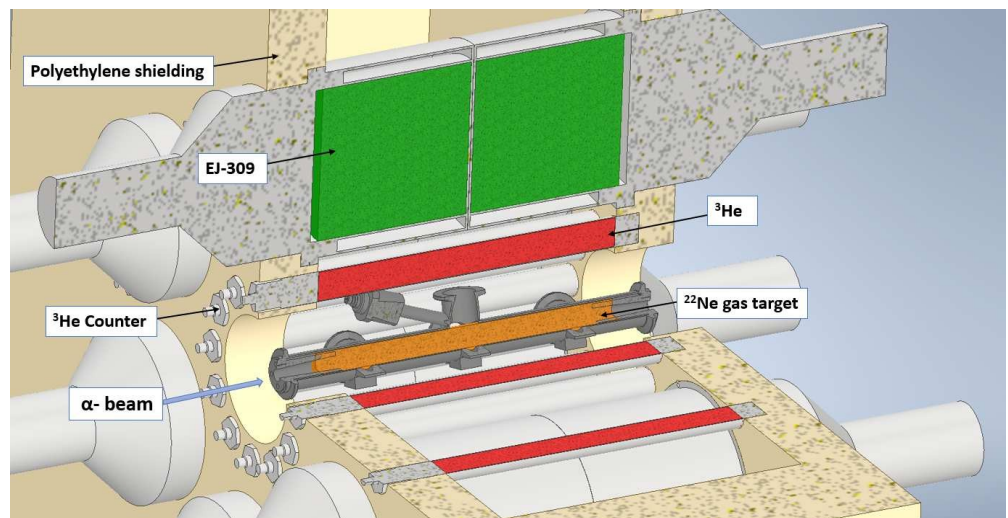

**Figure 12.** Cutoff view of the SHADES detector array surrounding the target chamber.

## 6. Conclusions

In the framework of the multi-messenger Astronomy era, we are collecting more and more precise data from the Universe.

Such accurate observations require a stellar model to rely on high-precision inputs. The cross-section of thermonuclear reactions taking place in stars and responsible for the production of most of the elements in the Universe is one of the main inputs of stellar models. At energies of interest, however, the reaction cross-sections are extremely small, making their measurement challenging. In this context, deep underground laboratories, pioneered by LUNA at LNGS, offer unique conditions for the measurement of key reactions in nuclear astrophysics, thanks to their dramatic reduction in the background. Such a reduction, however, might not be enough and to achieve the required sensitivity the setup, for example, the detection system and the shielding, must be designed with special care

We reported the latest designs of the detection-shielding setups used at LUNA and their role in recent challenging experimental studies.

The present work reviewed γ, particle and neutron detectors covering most of the recent LUNA scientific cases.

LUNA choices for the detection-shielding setup are the key to recent successful experimental campaigns and are the state of the art in nuclear astrophysics cross-section measurement, both concerning the achieved precision, as for the case of the $^2$H(p, γ)$^3$He reaction, and the achievable sensitivity, for example, for the case of the $^{17}$O+p reactions.

Future new detection system designs, which will push the sensitivity to new limits for experimental underground nuclear astrophysics and will allow us to assess key reactions for the chemical evolution of the Universe, have been presented in the framework of the scientific program of LUNA.

**Author Contributions:** All the authors contributed to the data collection and/or analysis of the experiments cited in the text. All authors have read and agreed to the published version of the manuscript.

**Funding:** This research was funded by INFN with contributions by other institutions as detailed in the acknowledgments section.

**Data Availability Statement:** No new data were created or analyzed in this study. Data sharing is not applicable to this article.

**Acknowledgments:** Support from the National Research, Development and Innovation Office NKFIH, Hungary (contract number PD129060) is acknowledged. C. A. and D.M. acknowledge funding from the European Research Council (ERC-StG 2019 #852016). L.B., J.M. and D.Ro. gratefully acknowledge support by the ELDAR (burning questions on the origin of the Elements in the Lives and Deaths of stARs) UKRI ERC StG (EP/X019381/1). Authors acknowledge ChETEC-INFRA (EU project #101008324).

**Conflicts of Interest:** The authors declare no conflicts of interest.